\documentclass[epj]{svjour}

\usepackage{graphics}
\usepackage{bm}
\usepackage{hyperref}

\begin{document}

\title{Influence of $s_{\pm}$ symmetry on unconventional superconductivity in pnictides above the Pauli limit -- two-band model study}
\titlerunning{Influence of $s_{\pm}$ symmetry on unconventional superconductivity in pnictides}
\author{Andrzej Ptok}
\mail{aptok@mmj.pl}
\institute{Institute of Physics, University of Silesia, 40-007 Katowice, Poland}
\date{Received: date / Revised version: date}
\abstract{
The theoretical analysis of the Cooper pair susceptibility shows the two-band Fe-based superconductors (FeSC) to support the existence of the phase with nonzero Cooper pair momentum (called the Fulde--Ferrel--Larkin--Ovchinnikov phase or shortly FFLO), regardless of the order parameter symmetry. Moreover this phase for the FeSC model with $s_{\pm}$ symmetry is the ground state of the system near the Pauli limit. This article discusses the phase diagram $h-T$ for FeSC in the two-band model and its physical consequences. We compare the results for the superconducting order parameter with {\it s-wave} and $s_{\pm}$-{\it wave} symmetry -- in first case the FFLO phase can occur in both bands, while in second case only in one band. We analyze the resulting order parameter in real space -- showing that the FeSC with $s_{\pm}$-{\it wave} symmetry in the Pauli limit have typical properties of one-band systems, such as oscillations of the order parameter in real space with constant amplitude, whereas with {\it s-wave} symmetry the oscillations have an amplitude modulation. Discussing the free energy in the superconducting state we show that in absence of orbital effects, the phase transition from the BCS to the FFLO state is always first order, whereas from the FFLO phase to normal state is second order.
}

\maketitle

The final publication is available at {\bf link.springer.com}

\section{Introduction}

In type-II superconductors, the magnetic field destroys the  Bardeen--Cooper--Schrieffer (BCS) superconductivity in two competitive ways -- through orbital (diamagnetic) or paramagnetic pair-breaking effects. The first one is related to the Abrikosov vortex state, and destroys the superconductivity in external magnetic field $H_{c2}^{orb}$ when the vortex cores begins to overlap. The second one originates from the Zeeman splitting of single electron energy levels. The superconductivity is destroyed when the magnetic field reaches the critical value $ H_{c2}^{P} $  due to polarization of the electrons. The relative intensity of the effects would be described by the Maki parameter $\alpha = \sqrt{2} H_{c2}^{orb} / H_{c2}^{P}$.~\cite{maki.66} Usually the critical field necessary for the the orbital effects to destroy superconductivity is lower than the one required by the diamagnetic effect ($\alpha \ll 1$). However, for some materials  (e.g. heavy fermions systems), an increasing magnetic field would destroy superconductivity through paramagnetic effects ($\alpha \geq 1$). In this case, in high magnetic field (greater than $H_{c2}^{P}$) Cooper pairs may be formed with non-zero total momentum between Zeeman-split parts of the Fermi surface. This gives rise to oscillations of the superconducting order parameter in the real space. This phase is called the Fulde--Ferrel--Larkin--Ovchinnikov (FFLO) phase.~\cite{FF,LO}

There are strong indications that the FFLO phase can be observed in heavy fermions systems,~\cite{yin.maki.93,shimahara.94,casalbuoni.nardulli.04,adachi.ikeda.03,matsuda.shimahara.07} e.g. $CeCoIn_5$ which is a strong candidate for exhibiting this state.~\cite{bianchi.movshovich.03,bianchi.movshovich.02,tayama.harita.02,capan.bianchi.04,kayuyanagi.saitoh.05} It is a clean system,~\cite{movshovich.jaime.01} with a layered structure suggesting a quasi-2D nature of electrons~\cite{hall.palm.01} and the Maki parameter is estimated to be $\alpha \simeq 5$.~\cite{kumagai.saitoh.06} Moreover, theoretical works suggest that this phase can exist in the presence of impurities~\cite{agterberg.yang.01,wang.hu.07,cui.yang.08,ptok.10,ikeda.10} and incommensurate spin density waves,~\cite{ptok.maska.11,mierzejewski.ptok.10} which is consistent with the experimental results.~\cite{takiwa.movshovich.08,movshovich.tokiwa.09,tokiwa.movshovich.10,kenzelmann.strassle.08,kenzelmann.gerber.10}
  Because Fe-based superconductors (FeSC) also have some of these features (they are also layered~\cite{singh.du.08,ding.richard.08,kondo.santander.08,cvetkovic.tesanovic.2.09,cvetkovic.tesanovic.09} clean~\cite{kim.tanatar.11,khim.lee.11} materials with relatively high Maki parameter $\alpha \sim 1-2$ ~\cite{khim.lee.11,cho.kim.11,zhang.liao.11,kurita.kitagawa.11,terashima.kihou.13}) we can expect the existence of the FFLO phase.~\cite{gurevich.10,gurevich.11,ptok.crivelli.13}

The Fermi surfaces (FS) in FeSC are composed of hole-like Fermi pockets (around the $\Gamma = ( 0,0 )$ point) and electron-like Fermi pockets (around the $M = ( \pi , 0 )$ or $ ( 0 , \pi )$ point). By the analysis of the Cooper pair susceptibility in two-band model of FeSC, such systems are shown to support the existence of a FFLO phase, regardless of the exhibited order parameter (OP) symmetry.~\cite{ptok.crivelli.13} However theoretical results point to the presence of $s_{\pm} \sim \cos ( k_{x} ) \cdot \cos ( k_{y} )$ pairing symmetry in FeSC.
~\cite{mazin.singh.08,kuroki.onari.08,chubukov.efremov.08,seo.bernevig.08,parish.hu.08,gao.su.10,bang.choi.08,tasi.zhang.09,chubukov.vavilov.09,mazin.10,hirschfeld.horshunov.11,jang.hong.12} In this case the OP exhibits a sign reversal between the two FS sheets. It should be noted the state with nonzero Cooper pair momentum, in superconducting FeSC with $s_{\pm}$ symmetry, is the ground state of the system near the Pauli limit.~\cite{ptok.crivelli.13} Because of this, it seems reasonable to determine the influence of $s_{\pm}$ and $s$ symmetry on the physical properties of the FFLO phase in multi-band systems.

In this paper we consider a minimal two-band model of FeSC to study physical properties of the FFLO phase, such as the $h-T$ diagram for the superconducting state with {\it s-wave} and $s_{\pm}$-{\it wave} symmetry and relevant phase transitions, the different momentum of the Cooper pairs in the bands and their influence on the OP in real space.

\section{Two band model of iron-base superconductors and theoretical method}

\begin{figure}[!t]
\begin{center}
\includegraphics{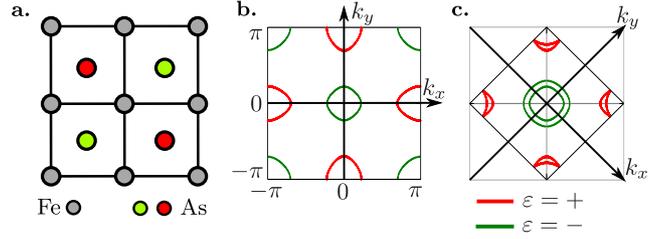} 
\caption{(Color on-line) (Panel a) $FeAs$ layer in pnictides. $Fe$ (dark dot) and $As$ (green and red dot) ions form a quadratic lattice. $As$ ions are placed above (red dot) or under (green dot) the centers of the squares formed by $Fe$. (Panel b) Fermi surface in effective and (Panel c) true Brillouin zone for $\mu = 1.54 | t_{1} |$ in minimal two-band models describing iron-base superconductors proposed by Ref.~\cite{raghu.qi.08}.}
\label{fig.bandgesc}
\end{center}
\end{figure}

In this part we set up the FeSC system using a minimal two-orbital per site model, with hybridization between the $d_{xz}$ and $d_{yz}$ orbitals. We adopt the band structure proposed in Ref.~\cite{raghu.qi.08}. The Hamiltonian takes the form:
\begin{eqnarray}\label{eq.hamC}
H_{0} &=& \sum_{{\bm k}\sigma} \sum_{\alpha\beta} T^{\alpha\beta}_{{\bm k}\sigma} c_{\alpha{\bm k}\sigma}^{\dagger} c_{\beta{\bm k}\sigma}
\end{eqnarray}
where $c_{\alpha{\bm k}\sigma}^{\dagger}$ ($c_{\alpha{\bm k}\sigma}$) is the creation (annihilation) operator of particles with momentum ${\bm k}$ and spin $\sigma$ in the orbital~$\alpha$. $T^{\alpha\beta}_{{\bm k}\sigma} = T^{\alpha\beta}_{\bm k} - ( \mu + \sigma h ) \delta_{\alpha\beta}$ is the kinetic energy term of a particle with momentum ${\bm k}$ changing the orbital from $\beta$ to $\alpha$ and is given by:
\begin{eqnarray}
\nonumber T^{11}_{\bm k} &=& - 2 \left( t_{1} \cos ( k_{x} ) + t_{2} \cos ( k_{y} ) \right) - 4 t_{3} \cos ( k_{x} ) \cos ( k_{y} ) , \\ 
\nonumber T^{22}_{\bm k} &=& - 2\left( t_{2} \cos ( k_{x} ) + t_{1} \cos ( k_{y} ) \right) - 4 t_{3} \cos ( k_{x} ) \cos ( k_{y} ) ,  \\ 
T^{12}_{\bm k} &=& T^{21}_{\bm k} = - 4 t_{4} \sin ( k_{x} ) \sin ( k_{y} ) . 
\end{eqnarray}
The hoppings have values: $(t_{1},t_{2},t_{3},t_{4}) = (-1.0,1.3,-0.85,-0.85)$, in units of $| t_{1} |$. $\mu$ is the chemical potential and $h$ is the external magnetic field parallel to the lattice plane, allowing us to neglect orbital effects . At half-filling, a two electrons per site configuration requires $\mu = 1.54 | t_{1} |$. In this case we have two FSs -- giving an electron-like band ($\varepsilon = +$) and hole-like band ($\varepsilon=-$) -- Fig.~\ref{fig.bandgesc}.b.

By diagonalizing the Hamiltonian (\ref{eq.hamC}), one obtains:
\begin{eqnarray}\label{eq.ham.mom}
H'_{0} &=& \sum_{\varepsilon{\bm k}\sigma} E_{\varepsilon{\bm k}\sigma} d_{\varepsilon{\bm k}\sigma}^{\dagger} d_{\varepsilon{\bm k}\sigma}
\end{eqnarray}
with eigenvalues $E_{\varepsilon{\bm k}\sigma} = E_{\varepsilon{\bm k}} - ( \mu + \sigma h )$, where:
\begin{equation}
E_{\pm,{\bm k}} = \frac{ T_{\bm k}^{11} + T_{\bm k}^{22} }{2} \pm \sqrt{ \left( \frac{ T_{\bm k}^{11} - T_{\bm k}^{22} }{2} \right)^{2} + \left( T_{\bm k}^{12} \right)^{2} } ,
\end{equation}
$d_{\varepsilon{\bm k}\sigma}^{\dagger}$ is a new fermion quasi-particle operator in the band $\varepsilon = \pm$.

Here it should be noted that present model is insufficient to approximate the full band structure, especially with regard to the correct orbital weights along the FS sheets. Real pnictides $FeAs$ layers are built by $Fe$ ions forming a square lattice surrounded by $As$ ions which also form a square lattice (Fig.~\ref{fig.bandgesc}.a).~\cite{kamihara.waranabe.08,rotter.tegel.08,stewart.11,kordyuk.12,hoffman.11,sadovskii.08} $As$ ions are placed above or under the centers of the squares formed by $Fe$. This leads to two inequivalent positions of $Fe$ atoms, so that there are two ions of $Fe$ and $As$ in an elementary cell. If the primitive unit cell is taken to be a square containing a single $Fe$ atom, the effective Brillouin zone (BZ) is the square shown in Fig.~\ref{fig.bandgesc}.b. The true primitive unit cell contains two $Fe$ ions -- the true BZ is twice as small. The FS around $(0,0)$ and $(\pi,\pi)$ are hole pockets asso
 ciated with $E_{-,{\bm k} \sigma} = 0$ and the FS around $(\pi,0)$ and $(0,\pi)$ are electron pockets from $E_{+{\bm k} \sigma} = 0$. The FS around $(\pi,\pi)$ is an artifact of the two-orbital approximation. The true BZ can obtained by folding the effective BZ -- the result of such downfolding is given in Fig.~\ref{fig.bandgesc}.c.~\cite{sadovskii.08,dagotto.moreo.12,graser.maier.09} It is evident that the FS obtained in this way are in qualitative agreement with the result of LDA calculations (only the third, less relevant, small hole-like pocket at the center is absent).~\cite{singh.du.08,ding.richard.08,kondo.santander.08,lioa.kondoa.09,cvetkovic.tesanovic.2.09,cvetkovic.tesanovic.09} The comparison between two- and more-bands model can be found e.g. in Ref.~\cite{dagotto.moreo.12,graser.maier.09}.

In the orbital basis $C_{{\bm k}\sigma} = ( c_{1{\bm k}\sigma} , c_{2{\bm k}\sigma} )^{T}$ we are able to consider interactions such as the Hubbard repulsion for electrons in the same orbital, Hubbard-like repulsion between different orbitals, ferromagnetic Hund coupling and a pair-hopping term.
~\cite{graser.maier.09,raghuvanshi.singh.11,daghofer.moreo.08,kubo.07,manousaki.ren.08} However in the band basis $D_{{\bm k}\sigma} = ( d_{+,{\bm k}\sigma} , d_{-,{\bm k}\sigma} )^{T}$ an effective superconducting pairing can be set between only the quasi-particles inside each band ~\cite{seo.bernevig.08,parish.hu.08,gao.su.10,sykora.coleman.11,korshunov.togushova.13,masuda.kurihara.10,vorontsov.vavilov.09,lu.zou.09,huang.gao.12}, if the intraband pairing interaction dominates~\cite{hirschfeld.horshunov.11}.
In this case the superconductivity in the FFLO phase, assuming only one momentum in each band, can be effectively expressed by the Hamiltonian:
\begin{eqnarray}\label{eq.Hsc}
H'_{SC} &=& \sum_{\varepsilon{\bm k}} \left( \Delta_{\varepsilon{\bm k}} d_{\varepsilon{\bm k}\uparrow}^{\dagger} d_{\varepsilon,-{\bm k}+{\bm q}_{\varepsilon} \downarrow}^{\dagger} + H.c. \right ) ,
\end{eqnarray}
where $\Delta_{\varepsilon{\bm k}} = \Delta_{\varepsilon} \eta ( {\bm k} )$ is the amplitude of the OP for Cooper pairs with total momentum ${\bm q}_{\varepsilon}$ (in band $\varepsilon$ with symmetry described by $\eta( {\bm k} )$). The structure factor is given by  $\eta ( {\bm k} ) = 1$ for {\it s-wave} and  $\eta ( {\bm k} ) = 4 \cos( k_{x} ) \cos ( k_{y} )$ for $s_{\pm}$-{\it wave} symmetry of the OP.~\cite{ptok.crivelli.13}

Using the Bogoliubov transformation we can find a final fermion basis $\Gamma_{\varepsilon{\bm k}} = ( \gamma_{\varepsilon{\bm k}\uparrow} , \gamma_{\varepsilon,-{\bm k}\downarrow} )^{T}$, describing the quasi-particle excitations in the superconducting state:
\begin{eqnarray}\label{eq.hamG}
H = \sum_{\varepsilon{\bm k}\tau} \bar{E}_{\varepsilon{\bm k}\tau} \gamma_{\varepsilon{\bm k}\tau}^{\dagger} \gamma_{\varepsilon{\bm k}\tau} + const.
\end{eqnarray}
with
\begin{eqnarray}
\bar{E}_{\varepsilon{\bm k}\tau} &=& \frac{ E_{\varepsilon{\bm k}\uparrow} - E_{\varepsilon,-{\bm k}+{\bm q}_{\varepsilon}\downarrow} }{2} \\
\nonumber &+& \tau \sqrt{ \left( \frac{ E_{\varepsilon{\bm k}\uparrow} + E_{\varepsilon,-{\bm k}+{\bm q}_{\varepsilon}\downarrow} }{2} \right)^{2} + | \Delta_{\varepsilon{\bm k}} |^{2} } 
\end{eqnarray}
where $\tau = \pm$. The free energy is given by:
\begin{eqnarray}\label{eq.enesw}
\Omega_{\varepsilon} ( {\bm q}_{\varepsilon} , \Delta_{\varepsilon} ) &=& -k T \sum_{{\bm k}\tau} \ln \left( 1 + \exp ( - \beta \bar{E}_{\varepsilon{\bm k}\tau} ) \right) \\
\nonumber &+& \sum_{\bm k} \left( E_{\varepsilon{\bm k}\downarrow} - \frac{ \gamma | \Delta_{\varepsilon{\bm k}} |^{2} }{ V_{\varepsilon} } \right) ,
\end{eqnarray}
where $V_{\varepsilon}$ is the interaction intensity in band $\varepsilon$ ($\gamma = 1$ for {\it s-wave} and $\gamma = 2$ for $s_{\pm}$-{\it wave} symmetry). The ground state in band for fixed $h$ and $T$ is found by minimizing the free energy w.r.t. the OPs and momentum ${\bm q}_{\varepsilon}$.

The effective description of the superconducting state is relatively simple, as a two-band model with gaps $\Delta_{\varepsilon{\bm k}}$ and normal-state dispersion $E_{\varepsilon{\bm k}\sigma}$.  We can see the advantages of this method when we rewrite Eq.~(\ref{eq.Hsc}) in original fermion basis $C_{{\bm k}\sigma}$. The transformation from the orbital-basis $C_{{\bm k}\sigma}$ to the band-basis $D_{{\bm k}\sigma}$ is given by:
\begin{eqnarray}\label{eq.trans}
D_{{\bm k}\sigma} = P_{\bm k} C_{{\bm k}\sigma} 
\end{eqnarray}
where $P_{\bm k}$ is the transformation matrix:
\begin{eqnarray}
P_{\bm k} &=& \frac{1}{\sqrt{1 + \zeta_{\bm k}^{2} }} \times \left( \begin{array}{cc}
1 & \zeta_{\bm k} \\ 
-\zeta_{\bm k} & 1
\end{array} \right) , \\
\nonumber \zeta_{\bm k} &=& \frac{ T^{12}_{\bm k} }{ \left( T_{\bm k}^{11} - T_{\bm k}^{22} \right) / 2  + \sqrt{ \left( T_{\bm k}^{11} - T_{\bm k}^{22} \right)^{2} / 4 + \left( T_{\bm k}^{12} \right)^{2} } } . \\ \label{eq.formfactor}
\end{eqnarray} 
In the original basis ($C_{{\bm k}\sigma}$) the Hamiltonian (\ref{eq.Hsc}) can be rewritten as:
\begin{eqnarray}
H_{SC} &=& \sum_{\alpha\beta\varepsilon{\bm k}} \left( \Delta^{\alpha\beta}_{\varepsilon{\bm k}} c_{\alpha{\bm k}\uparrow}^{\dagger} c_{\beta,-{\bm k}+{\bm q}_{\varepsilon}\downarrow}^{\dagger}  + H.c. \right) ,
\end{eqnarray}
where $\Delta^{11}_{\varepsilon{\bm k}}$ and $\Delta^{22}_{\varepsilon{\bm k}}$ are the intra-orbital OPs while $\Delta^{12}_{\varepsilon{\bm k}}$ and $\Delta^{21}_{\varepsilon{\bm k}}$ are the inter-orbital OPs:
\begin{eqnarray}
\Delta^{11}_{\varepsilon{\bm k}} &=& \frac{ \Delta_{+,{\bm k}} \delta_{\varepsilon,+} + \Delta_{-,{\bm k}} \zeta_{\bm k} \zeta_{-{\bm k}+{\bm q}_{\varepsilon}} \delta_{\varepsilon,-} }{ \sqrt{ 1 + \zeta_{\bm k}^{2} } \sqrt{ 1 + \zeta_{-{\bm k}+{\bm q}_{\varepsilon}}^{2} } } , \label{eq.op11}
\\
\Delta^{12}_{\varepsilon{\bm k}} &=& \frac{ \Delta_{+,{\bm k}} \zeta_{-{\bm k}+{\bm q}_{\varepsilon}} \delta_{\varepsilon,+} - \Delta_{-,{\bm k}} \zeta_{\bm k} \delta_{\varepsilon,-} }{ \sqrt{ 1 + \zeta_{\bm k}^{2} } \sqrt{ 1 + \zeta_{-{\bm k}+{\bm q}_{\varepsilon}}^{2} } } , \label{eq.op12}
\\
\Delta^{21}_{\varepsilon{\bm k}} &=& \frac{ \Delta_{+,{\bm k}} \zeta_{\bm k} \delta_{\varepsilon,+} - \Delta_{-,{\bm k}} \zeta_{-{\bm k}+{\bm q}_{\varepsilon}} \delta_{\varepsilon,-} }{ \sqrt{ 1 + \zeta_{\bm k}^{2} } \sqrt{ 1 + \zeta_{-{\bm k}+{\bm q}_{\varepsilon}}^{2} } } , \label{eq.op21}
\\
\Delta^{22}_{\varepsilon{\bm k}} &=& \frac{ \Delta_{+,{\bm k}} \zeta_{\bm k} \zeta_{-{\bm k}+{\bm q}_{\varepsilon}} \delta_{\varepsilon,+} + \Delta_{-,{\bm k}} \delta_{\varepsilon,-} }{ \sqrt{ 1 + \zeta_{\bm k}^{2} } \sqrt{ 1 + \zeta_{-{\bm k}+{\bm q}_{\varepsilon}}^{2} } } , \label{eq.op22}
\end{eqnarray}
while in real space it can be written as:
\begin{equation}
\label{eq.hamrec}
H_{SC} = \sum_{ij\alpha\beta} \left( \Delta_{ij}^{\alpha \beta} c_{\alpha i \uparrow}^{\dagger} c_{\beta j \downarrow}^{\dagger} + H.c. \right)
\end{equation}
where 
\begin{equation}
\label{eq.opre}
\Delta_{ij}^{\alpha \beta} = \frac{1}{N} \sum_{\varepsilon {\bm k}} \Delta_{\varepsilon{\bm k}}^{\alpha\beta} \exp ( - i {\bm k} \cdot {\bm R}_{i} ) \exp ( - i ( -{\bm k}+{\bm q}_{\varepsilon} ) \cdot {\bm R}_{j} ) .
\end{equation}
In the $D_{{\bm k}\sigma}$ basis we have formally a two band system with two independent bands $\varepsilon = \pm$. However the effective description (by $H'_{0} + H'_{SC}$) of the system, corresponds to a full description (by $H_{0} + H_{SC}$) with interactions between electrons with opposite spins on sites $i$ (in orbital $\alpha$) and $j$ (in orbital $\beta$) of the lattice (intra-orbital $\alpha = \beta$ and also inter-orbital $\alpha \neq \beta$ pairing in real space -- Eq.~(\ref{eq.op11})-(\ref{eq.op22})). The transformation given by Eq.~\ref{eq.trans} can be treated as a mapping from the orbital basis $C_{{\bm k}\sigma}$ to the band basis $D_{{\bm k}\sigma}$ in which the Hamiltonian is tridiagonal ~\cite{vorontsov.vavilov.09,lu.zou.09}, which is exactly diagonalized by the Bogoliubov transformation to Eq.~(\ref{eq.hamG}).~\cite{sykora.coleman.11}

\section{Numerical results and discussion}

Consistently with the calculations presented we formally consider the two bands $\varepsilon = \pm$ as independent.  In this case, in each of these bands there may be another effective pairing potential $V_{\varepsilon}$, which allows for a different value of the amplitude $\Delta_{\varepsilon}$ and values of the critical parameters (e.g. $h_{C\varepsilon}$) in each band. However, experimental results show the OP in both bands to vanish at the same critical temperature.~\cite{ding.richard.08} Accordingly we adopted $V_{+} = -2.69 | t_{1} |$ and $V_{-} = - 7.7 | t_{1} |$ for {\it s-wave} and $V_{+} = -0.605 | t_{1} |$ and $V_{-} = - 1.44 | t_{1} |$ for $s_{\pm}$-{\it wave}. Then $\Delta_{+} = \Delta_{-} = 0.2 | t_{1} |$ for {\it s-wave} and $\Delta_{+} = \Delta_{-} = 0.1 | t_{1} |$ for $s_{\pm}$-{\it wave} for $h = 0 | t_{1} |$ and $k T = 10^{-4} | t_{1} |$. All numerical computations were carried out for a square lattice $N_{X} \times N_{Y} = 1200 \times 1200$ with periodic b
 oundary conditions.

\begin{figure}[!t]
\begin{center}
\includegraphics{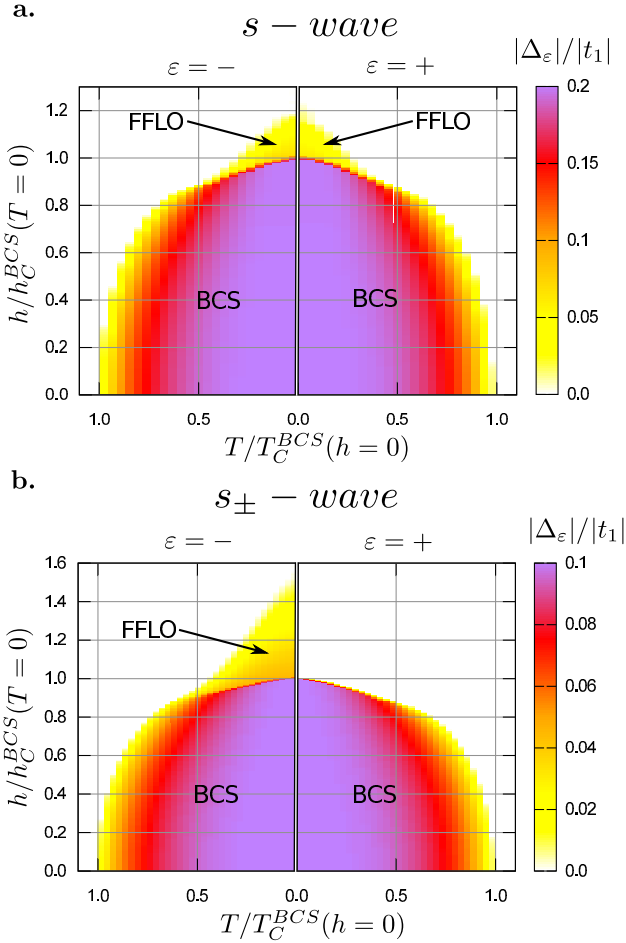}
\caption{(Color on-line) $h-T$ phase diagram. Color intensity is proportional to the amplitude $| \Delta_{\varepsilon} |$. The results are for {\it s-wave} and $s_{\pm}$-{\it wave} symmetry of the order parameter in each band.}
\label{fig.ht}
\end{center}
\end{figure}

\subsection{h-T phase diagram and phase transition}

The phase diagram $h-T$ (Fig.~\ref{fig.ht}) was determined minimizing the free energy $\Omega_{\varepsilon}$ w.r.t. $\Delta_{\varepsilon}$ and ${\bm q}_{\varepsilon}$. With our choice of parameters, the BCS phase disappears in both bands at the same magnetic field $h_{C+}^{BCS} ( T ) \simeq h_{C-}^{ BCS} ( T )$. For both analyzed symmetries, we can see three phase transitions: the transition from the BCS phase to the normal phase or the FFLO phase (in magnetic field $h_{C \varepsilon}^{BCS}$) and the transition from the FFLO phase to the normal state (in magnetic field $h_{C \varepsilon}^{FFLO}$). Additionally the FFLO phase can occur below a characteristic temperature $T^{\ast} \simeq 0.48 T_{C}^{BCS} ( h = 0 ) $. At $h_{C \varepsilon}^{BCS}$ and above $T^{\ast}$ we observe the second phase transition. Below $T^{\ast}$ this transition from the BCS phase to the FFLO becomes first order, but from FFLO phase to normal state it is still second order.
This is typical of the FFLO phase in one-band systems in absence of orbital effects, and was shown in a number of theoretical ~\cite{matsuda.shimahara.07,adachi.ikeda.03,agterberg.yang.01,ptok.maska.09,houzet.buzdin.01,houzet.mineev.06,yang.sondhi.98,maki.tsuneto.64,burkhardt.rainer.94} and experimental works.
~\cite{bianchi.movshovich.03,bianchi.movshovich.02,tayama.harita.02,capan.bianchi.04,kumagai.saitoh.06,izawa.yamaguchi.01,watanabe.kasahara.04} 
Moreover, in FeSC more than one phase transition can be experimentally observed in the FFLO phase regime -- at low temperature and high magnetic field (LTHM),~\cite{terashima.kihou.13,zocco.grube.13,burger.hardy.13} however there is a lack of clear evidence on the order of these transitions.

\begin{figure}[!t]
\begin{center}
\includegraphics{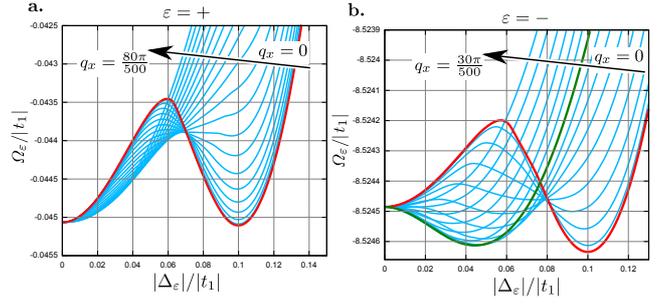}
\caption{(Color on-line) Free energy $\Omega_{\varepsilon} ( {\bm q}_{\varepsilon} , \Delta_{\varepsilon} ) / N$ in case of $s_{\pm}$ symmetry, for different ${\bm q}_{\varepsilon} = ( q_{x} , 0 )$ growing from the right to the left of the plot, near the critical magnetic field $h \simeq h_{C}^{BCS}$ and $kT = 10^{-4} |t_{1}|$. Red line corresponds to the BCS state ${\bm q}_{\varepsilon} = ( 0 , 0 )$ while the green line to the FFLO state with ${\bm q}_{-} = ( \frac{18 \pi}{500} , 0 )$, which is the ground state above $h_{C}^{BCS}$ in $\varepsilon = -$.}
\label{fig.freene}
\end{center}
\end{figure}

In the LTHM regime we observe the FFLO phase in bands $\varepsilon = \pm$ for {\it s-wave} symmetry (Fig.~\ref{fig.ht}.a), but only in band $\varepsilon = -$ for $s_{\pm}$-{\it wave} (Fig.~\ref{fig.ht}.b). Since we formally describe two independent bands, the phenomenological Ginzburg-Landau theory applies to each, allowing to write the free energy (for band $\varepsilon$ with momentum ${\bm q}_{\varepsilon}$) as a function of the amplitude of the OP $\Delta_{\varepsilon}$:~\cite{casalbuoni.nardulli.04}
\begin{eqnarray}
&& \delta \Omega_{\varepsilon} = \Omega_{\varepsilon} ( {\bm q}_{\varepsilon} , \Delta_{\varepsilon} ) - \Omega_{\varepsilon} ( {\bm q}_{\varepsilon} , 0 ) \\
\nonumber &\simeq& a_{1} | \Delta_{\varepsilon} |^{2} + a_{2} | \Delta_{\varepsilon} |^{4} + a_{3} | \Delta_{\varepsilon} |^{6} + a_{4} | \Delta_{\varepsilon} |^{8} + \mathcal{O} ( | \Delta_{\varepsilon} |^{10} ) ,
\end{eqnarray}
where the coefficients $a_{i}$ depend on the external magnetic field $h$ and temperature $T$. 
As we see in Fig.~\ref{fig.freene} ($\Omega_{\varepsilon}$ for $s_{\pm}$-{\it wave} symmetry), $a_{i}$ also strongly depend on $q_{\varepsilon}$ and type of band. The profile of the free energy for the BCS state (red lines) point to a first order transition near $h_{C}^{BCS}$ for the both bands.
On the other hand, the minimum energy landscape has a different character for states with non-zero momentum ${\bm q}_{\varepsilon}$. In the case of $\varepsilon = +$ (Fig.~\ref{fig.freene}.a) the global energy minimum $\Omega_{+}$ in a superconducting phase is always attained by a BCS state (for ${\bm q}_{+} = 0$ -- red dot in Fig.~\ref{fig.momentumcp}.d). In the second case for band $\varepsilon = -$ (Fig.~\ref{fig.freene}.b) the function $\Omega_{-}$ in an external magnetic field $h > h_{C}^{BCS}$ for ${\bm q}_{-} \sim ( \frac{20}{500} \pi , 0 )$ (FFLO phase) has a typical form with first order transition to the normal state (global minimum shown by the blue dot in Fig.~\ref{fig.momentumcp}.d). This case also occurs in both bands for {\it s-wave} symmetry (in fig~\ref{fig.momentumcp}.c).

For the BCS state the phase diagram takes its typical form for both symmetries. The initial slope of $h_{C}^{BCS} ( T )$ at $T_{C}$ is infinite and at temperature $T^{\ast}$ we should observe a discontinuity of $dh_{C}/dT$. This follows from the fact that the upper magnetic critical field above $T^{\ast}$ is equal to $h_{C \varepsilon}^{BCS} ( T )$, while below $T^{\ast}$ is equal to $h_{C \varepsilon}^{FFLO} ( T )$. In the present model we found $h_{C \varepsilon}^{FFLO} \simeq 1.22 h_{C \varepsilon}^{BCS}$ ($1.55 h_{C \varepsilon}^{BCS}$) for $s$ ($s_{\pm}$) symmetry at $T \rightarrow 0$. It should be noted that this result concerns the case of Cooper pairs having a single momentum ${\bm q}_{\varepsilon}$. The analysis for a larger number of allowed Cooper pairs momenta can lead to much higher values of $H_{c2}$.~\cite{LO,ptok.10,shimahara.98,shimahara.99,mora.combescot.04,mora.combescot.05} However, this does not qualitatively affect the results presented.

\subsection{Momentum of the Cooper pairs and order parameter in real space}

\begin{figure}
\begin{center}
\includegraphics{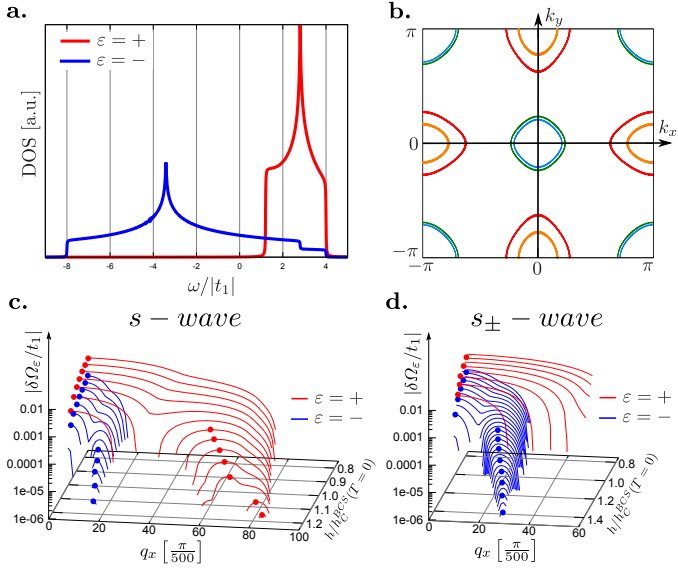}
\caption{(Color on-line) (Panel a) Density of states  for $\mu = 1.54 | t_{1} |$ for the discussed minimal two-band model in absence of external magnetic field. (Panel b) Splitting of the Fermi surface for electron with spin $\uparrow$ and $\downarrow$ in band $\varepsilon = +$ (red and orange line) and $\varepsilon = -$ (green and azure line) in presence of external magnetic field $h = 0.15 | t_{1} |$. (Panel c (resp. d)) Free energy difference $\delta \Omega_{\varepsilon}$ between the (superconducting) ground state and normal state in band $\varepsilon$ for given external magnetic field $h$ and momentum of Cooper pairs ${\bm q}_{\varepsilon} = ( q_{x} , 0 )$, for {\it s-wave} (resp. $s_{\pm}$-{\it wave}) symmetry. Red and blue dots show the place of the global ground state for the superconducting state.}
\label{fig.momentumcp}
\end{center}
\end{figure}

For {\it s-wave} symmetry, the Cooper pairs have greater momentum in band $\varepsilon = +$ than in band $\varepsilon = -$. The results are consistent with previous data obtained using the static Cooper pairs susceptibility.~\cite{ptok.crivelli.13} This is due to the construction of the bands in our model (Fig.~\ref{fig.momentumcp}.a). The external magnetic field causes a narrower splitting of the FS for wide bands, as band $\varepsilon = +$ ($\varepsilon = -$) has width $\delta E_{+} \approx 3 | t_{1} |$ ($\delta E_{-} \approx 12 | t_{1} |$). Thus, the critical magnetic field splits the band $\varepsilon = +$ more (Fig.~\ref{fig.momentumcp}.b). Numerical results indicate that the ground state corresponds to four equivalent momenta $\pm ( q_{\varepsilon} , 0 )$ and $\pm ( 0 , q_{\varepsilon} )$.~\cite{ptok.crivelli.13} This is also observed in one-band systems.~\cite{ptok.maska.09}

At the critical magnetic field $h_{C}^{BCS}$ Cooper pairs discontinuously acquire a non-zero total momentum ${\bm q}_{\varepsilon}$. Increasing the external magnetic field, the total momentum $| {\bm q}_{\varepsilon} |$ in the FFLO phase also increases while the amplitude $\Delta_{\varepsilon}$ decreases (Fig.~\ref{fig.momentumcp}.c and d), as reported also in the one-band systems.~\cite{ptok.maska.09}

Our proposed method to represent the FFLO phase in multi-band systems is an extension of the original method proposed by Linder and Sudb\o{} in Ref.~\cite{linder.sudbo.09}, used to describe the BCS phase in FeSC. They show that in case of the BCS phase (${\bm q}_{\varepsilon} = 0$), the inter-orbital pairing vanishes when $\Delta_{+} = \Delta_{-}$ and the OPs in both bands $\varepsilon$ have the same symmetry. This corresponds to a situation where only the intra-orbital pairing with symmetry $\eta ( {\bm k} )$ exists in the system. It effectively describes interactions between electrons on one (for {\it s-wave} symmetry) or two (for other symmetries like $s_{\pm}$, $d_{x^{2}-y^{2}}$, $d_{x^{2}y^{2}}$, etc.) sites of the lattice in real space. Also for our choice of $V_{\varepsilon}$, for both symmetries of the OP in the BCS state we have $\Delta_{+} = \Delta_{-}$. It is consistent with experimental data, in which the hole FS and electron FS pocket gaps are the same.~\cite{ding.richard.08,evtushinsky.inosov.09}. However in general this is not required. When $\Delta_{+} \neq \Delta_{-}$ a non-zero inter-band OP arises, but smaller in amplitude than the intra-band OP. The situation is more difficult in the FFLO phase case, when ${\bm q}_{+} \neq {\bm q}_{-} \neq 0$ and in both bands $\varepsilon$ the symmetry can be different, leading to $\Delta_{ij}^{\alpha\beta} \neq 0$ for any band and site of the lattice, and the {\it appearance} of a non-zero inter-band OP (Eq.~(\ref{eq.op12}) and (\ref{eq.op21})). Additionally the FFLO phase in either band is sufficient to break translational symmetry in real space.

For one band systems (or multi-band systems) with a FFLO phase (in one selected band), in which the Cooper pairs have one non-zero total momentum ${\bm q}$, the OP has constant amplitude in real space. The situation looks different in the case of multi-band systems, when Cooper pairs with different non-zero momenta ${\bm q}_{+} \neq {\bm q}_{-}$ exist. As we wrote, such systems can be described as a set of independent Bogoliubov bands $\varepsilon$ (in basis $D_{{\bm k}\sigma}$), in which the Cooper pairs can have total momentum ${\bm q}_{\varepsilon}$ and corresponding characteristic length scale $\zeta_{\varepsilon} \sim 1 / | {\bm q}_{\varepsilon} |$. However, in the original momentum space (with basis $C_{{\bm k}\sigma}$), this situation corresponds to different momenta of the Cooper pairs for intra- and inter-band paring (Eq.~(\ref{eq.op11})-(\ref{eq.op22})), and corresponding modulation amplitude of the OP in real space (from the non-equal characteristic length $\zeta_{
 +} \neq \zeta_{-}$.) This results are consistent with other theoretical works.~\cite{mizushima.takahashi.13} Numerical results obtained indicate that states with $s_{\pm}$-{\it wave} symmetry retain features of one-band systems.

\section{Summary}

The existence of $s_{\pm}$ symmetry in FeSC can have a measurable effect on the experimental results.~\cite{parish.hu.08,gao.su.10} In the Pauli limit it can be a stabilizing factor for unconventional superconductivity with non-zero total momentum of the Cooper pairs (the FFLO phase). We show that compared to {\it s-wave} symmetry, in FeSC with $s_{\pm}$-{\it wave} the FFLO phase occurs for a wider range on the $h-T$ phase diagram. Moreover in absence of orbital effects we determine the order of phase transitions for states with Cooper pairs with one momentum from the free energy -- the transition from the BCS phase to the FFLO phase is always first order, while it is second order from the FFLO phase to the normal state. States with one non-zero momentum of Cooper pairs are more stable in low temperature and high magnetic field than states with multiple-momenta in disordered systems.~\cite{agterberg.yang.01} However in absence of inhomogeneities, phases with linear combinations of Cooper pairs with different momentum can strongly affect the shape of the free energy and the order of the phase transitions.~\cite{mora.combescot.05}

In the considered model for $s_{\pm}$-{\it wave} symmetry, at low temperature and above the BCS critical magnetic field, the FFLO phase occurs only in one band leading to a phase with oscillating order parameter in real space with constant amplitude. Whereas for {\it s-wave} symmetry it is the preferred state in both bands, causing the order parameter to display an additional oscillation amplitude modulation. For this reason the FFLO phase in FeSC with $s_{\pm}$-{\it wave} symmetry has behavior analogous to one-band systems. The presented results are consistent with other theoretical works.~\cite{mizushima.takahashi.13}

\begin{acknowledgement}
I am grateful to Dawid Crivelli for insightful discussions, comments and help in preparing this paper.
\end{acknowledgement}

\end{document}